\documentclass[11pt]{article}
\usepackage{blois,epsfig}
\usepackage[latin1]{inputenc}
\usepackage[OT1]{fontenc}
\usepackage{psfrag}

\def\Journal#1#2#3#4{{#1} {\bf #2}, #3 (#4)}

\def\aap{\em A\&A}
\def\mnras{\em MNRAS}
\def\apj{\em ApJ}
\def\sci{\em Science}

\def\be{\begin{equation}}
\def\ee{\end{equation}}
\def\bea{\begin{eqnarray}}
\def\eea{\end{eqnarray}}

\begin{document}
\title{SSC SCENARIO FOR VHE EMISSION\\
  FROM 2 RADIOGALAXIES: M\,87 AND CEN\,A}

\author{ J.-P.~LENAIN, C.~BOISSON, H.~SOL }

\address{LUTH, Observatoire de Paris, CNRS, Universit\'e Paris Diderot,\\
5 Place Jules Janssen, 92190 Meudon, France}

\maketitle\abstracts{
  M\,87 is the first extragalactic source detected in the Very High Energy (VHE; $E > 100$\,GeV) $\gamma$-ray domain that is not a blazar, its large scale jet not being aligned to the line of sight. Slight modification of standard emission models of TeV blazars allows to account for the $\gamma$-ray spectra obtained with H.E.S.S. We present a multi-blob synchrotron self-Compton model taking explicitly into account large viewing angles and moderate values of the Lorentz factor as inferred from MHD simulations of jet formation. Predictions of the VHE emission for the nearby radiogalaxy Cen\,A and an interpretation of the broadband radiation of M\,87 are presented.
}

\section{The multi-blob SSC model and the VHE emission of M\,87 observed by H.E.S.S.}

The classic synchrotron self-Compton (SSC) scenarii cannot account for VHE emission for misaligned objects, and thus require further developments. We present here a modification of a previous model for blazars~\cite{2001A&A...367..809K,2003A&A...410..101K}, for an emitting zone close to the central supermassive black hole~\cite{2008A&A...478..111L}.

According to the results of general relativistic MHD simulations~\cite{2006MNRAS.368.1561M}, we can put constraints on the macrophysics parameters of our model, such as the distance of the Alfv\'en surface to the central black hole, the opening angle of the jet and the bulk Lorentz factor.

The emitting zone is represented by a 3D cap filled with several plasma blobs located at a distance $R_\mathrm{cap}$ from the black hole, just above the Alfv\'en surface to let time to the acceleration process to take place. The blobs are propagating in slightly different directions, enabling a differential Doppler boosting effect. We distinguish two extreme geometric cases in this framework:

\begin{itemize}
\item ``On-blob'': the line of sight is exactly aligned with the velocity vector of the central blob, leading to a maximal Doppler amplification.
\item ``Inter-blob'': the line of sight lies exactly through the gap between 3 adjacent blobs, leading to a minimal Doppler amplification.
\end{itemize}

\begin{figure}
  \begin{minipage}[c]{0.5\textwidth}
    \centering
    \psfig{figure=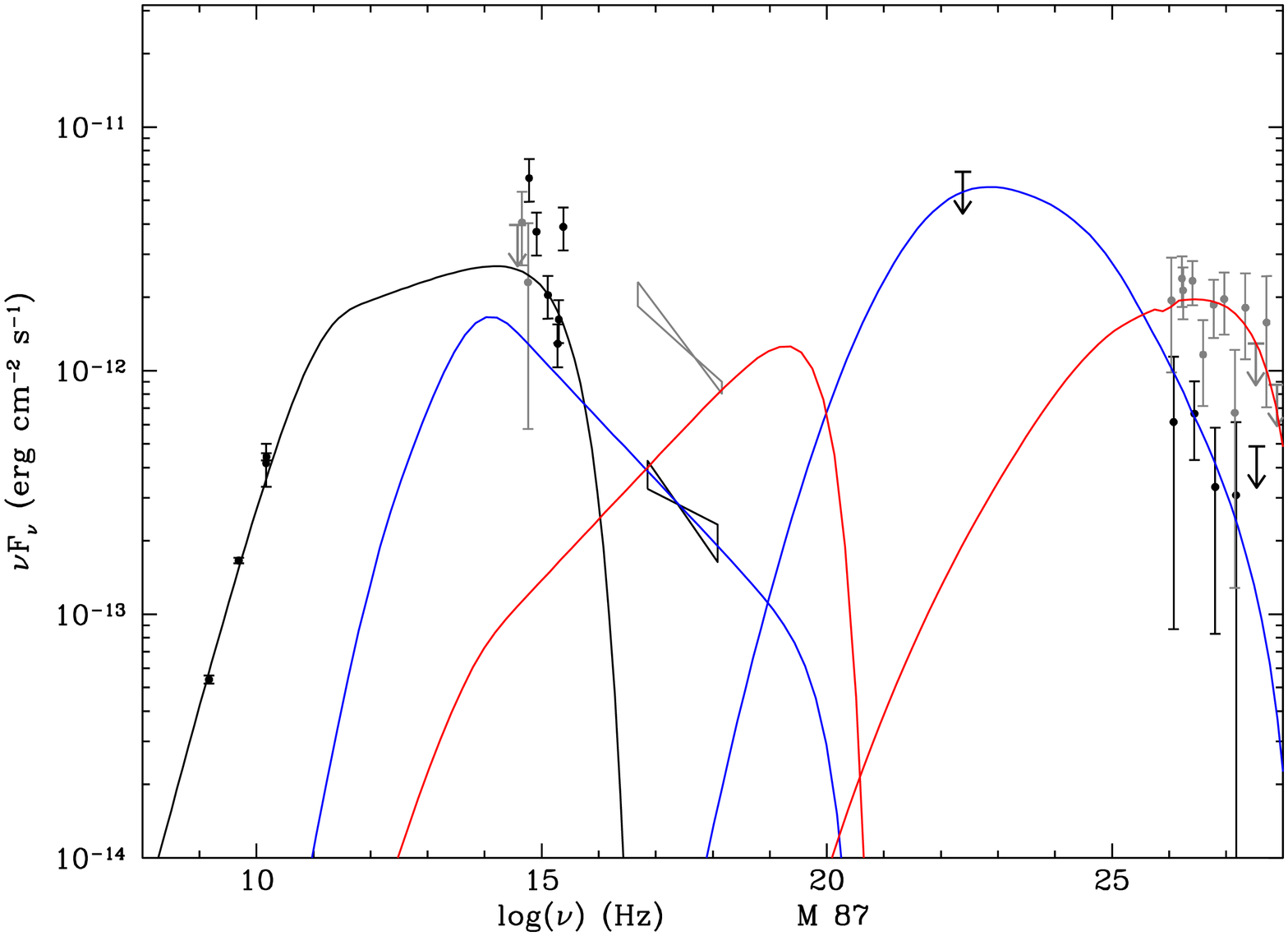,height=2.0in,angle=-90}
  \end{minipage}%
  \begin{minipage}[c]{0.5\textwidth}
    \centering
    \psfig{figure=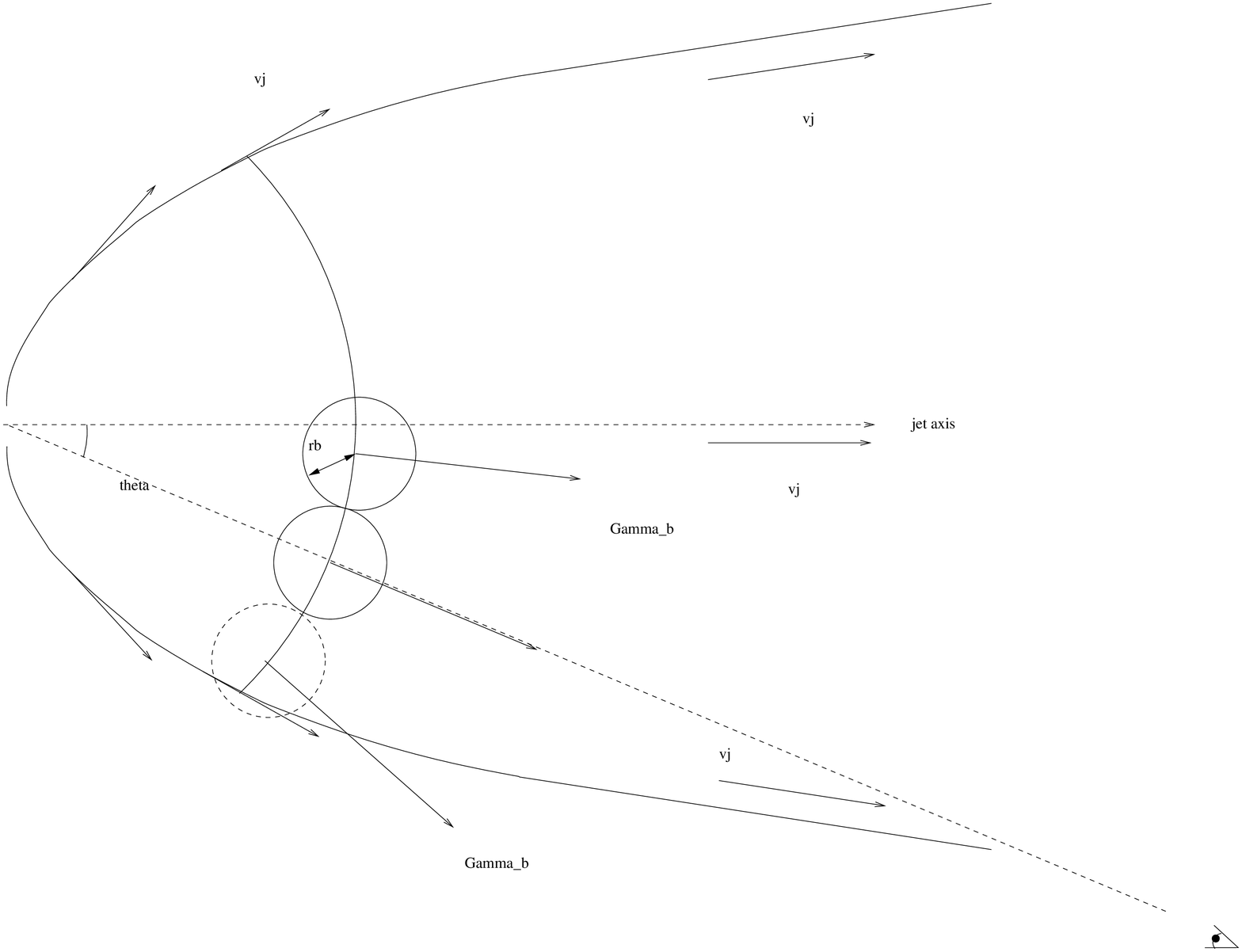,height=1.3in}
  \end{minipage}%
  \caption{{\it a)\/} SED of M\,87. The two states of activity observed in the VHE range can be reproduced within the multi-blob model, describing the H.E.S.S. data of 2004 in {\it blue lines} and 2005 in {\it red lines}. We predict a radical change of the X-ray regime between 2004 and 2005. {\it b)\/} Geometric scheme of the multi-blob model.
    \label{fig:M87_SED_Blois_multiblob}}
\end{figure}

We present in Fig.~\ref{fig:M87_SED_Blois_multiblob}a the results of our multi-blob model for the emission of the jet of M\,87, assuming that {\it Chandra} data~\cite{2003ApJ...599L..65P} taken in 2000 represent a low state of activity, comparable to the one observed by H.E.S.S.~\cite{2006Sci...314.1424A} in 2004.

\section{Predictions of VHE flux for Cen\,A}

Encouraged by the VHE detection of M\,87, we investigate a potential VHE emission from Cen\,A. This well-known radiogalaxy is 3.4\,Mpc away, even closer than M\,87 which lies at 16\,Mpc.

\begin{figure}
  \begin{minipage}[c]{0.5\textwidth}
    \centering
    \psfig{figure=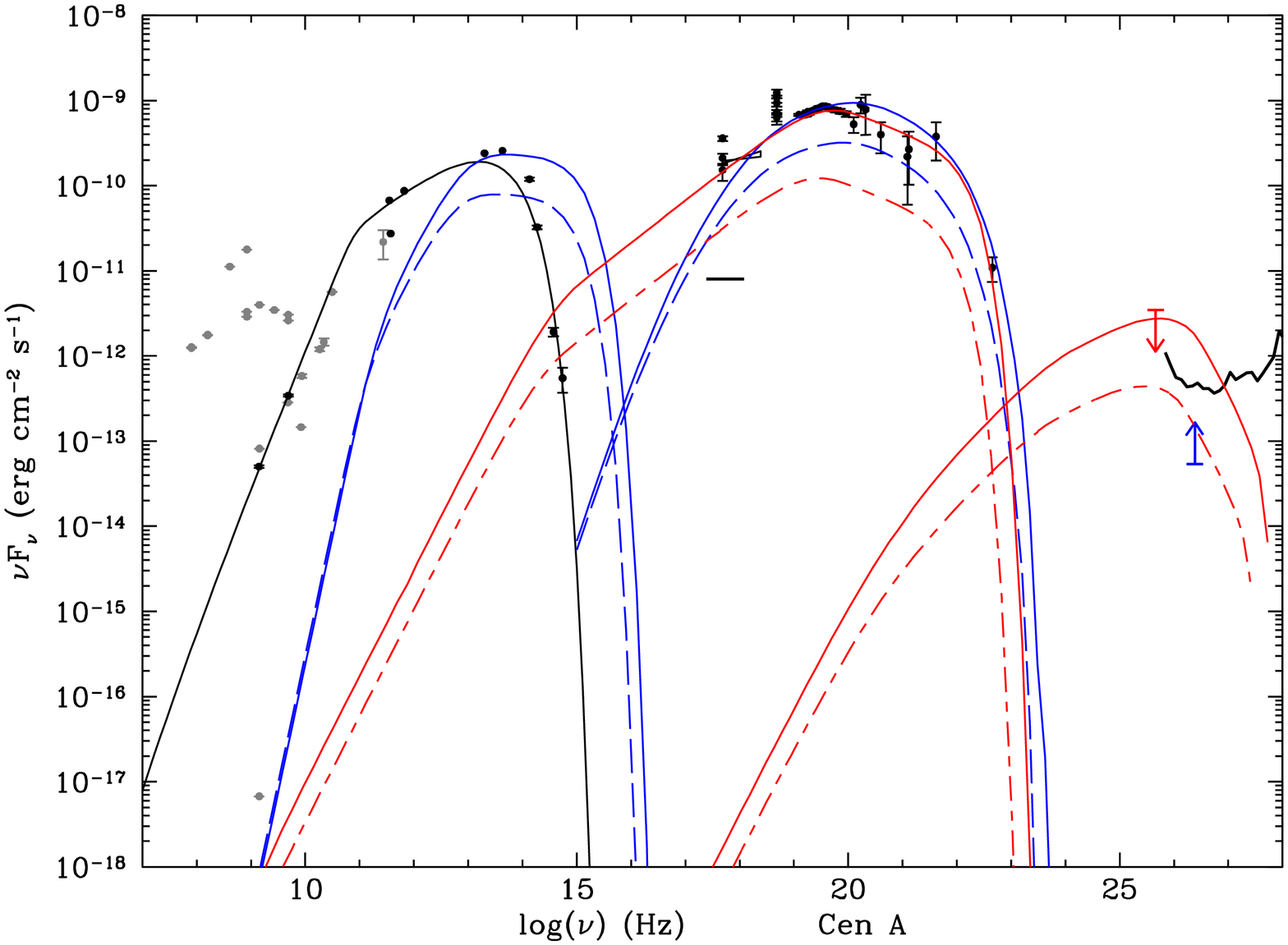,height=2.0in,angle=-90}
  \end{minipage}
  \begin{minipage}[c]{0.5\textwidth}
    \centering
    \psfig{figure=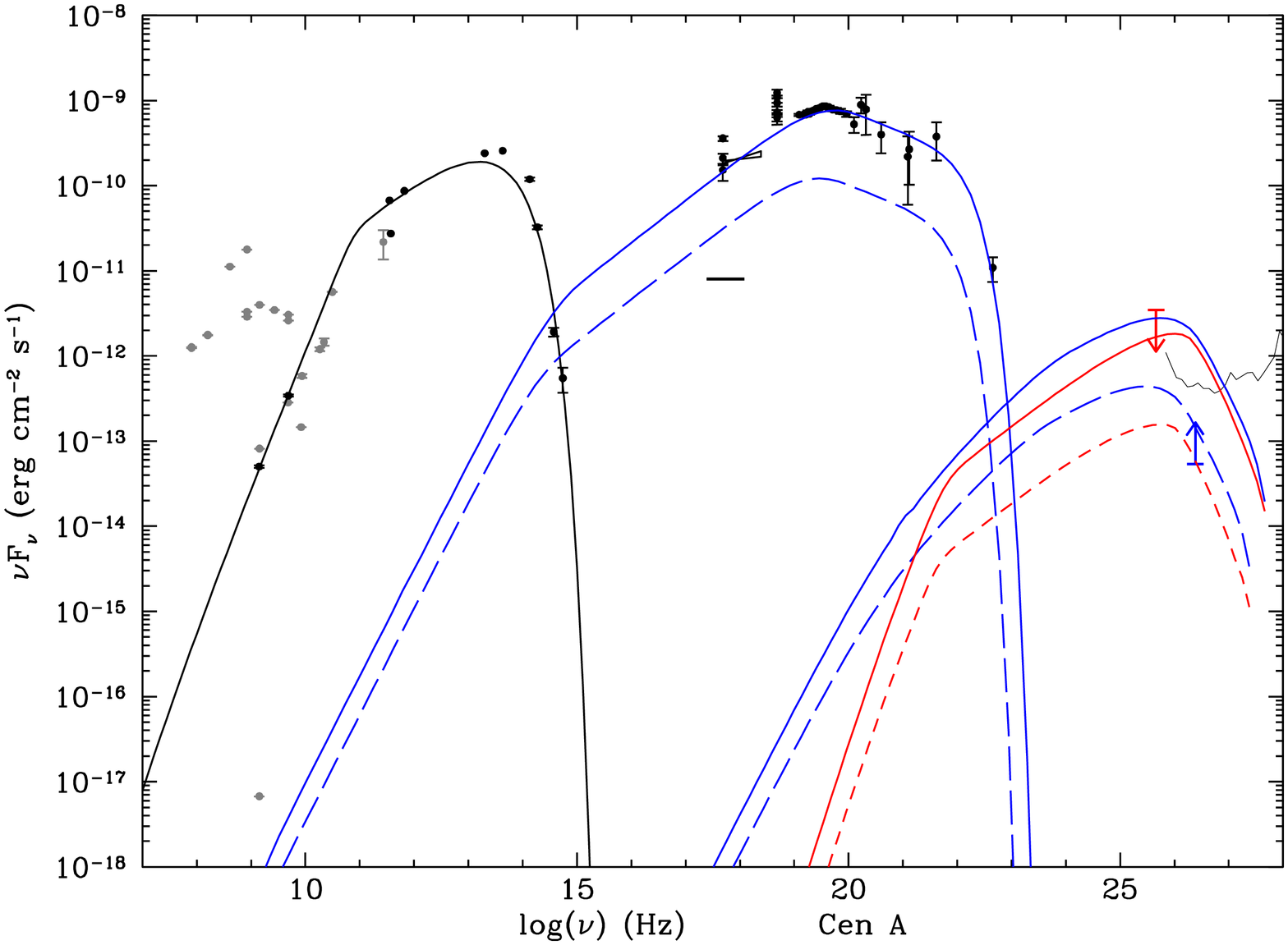,height=2.0in,angle=-90}
  \end{minipage}
  \caption{{\it a)\/} Two scenarii for the SED of Cen\,A within the multi-blob model, assuming the X- and soft $\gamma$-rays to be produced either by inverse Compton scattering ({\it blue lines}) or by synchrotron radiation ({\it red lines}). In the second case, detection is expected soon at TeV energies. {\it b)\/} SED of Cen\,A. The red line represents the external inverse Compton contribution of the mid-IR radiation field due to dust in the vicinity of the core of Cen\,A.
    \label{fig:CenA_SED}}
\end{figure}

We apply the multi-blob model to Cen\,A assumed here to be a misaligned blazar. Figure~\ref{fig:CenA_SED}a shows the spectral energy distribution (SED) of Cen\,A, with the VHE flux as expected from our multi-blob SSC model. At VHE frequencies, a typical sensitivity curve of H.E.S.S. is shown for a detection of 5$\sigma$ in 50\,h of observation at 30$^\circ$ of mean zenith angle ({\it V-shaped curve in black}) as well as the expected sensitivity limit of the future \v{C}erenkov Telescope Array (CTA, {\it blue lower limit}).

Recent mid-IR observations with the VLTI have revealed the presence of a dusty disk~\cite{2007A&A...471..453M} nearby the central engine of Cen\,A, with $T_\mathrm{dust} \sim 240$\,K. We show in Fig.~\ref{fig:CenA_SED}b the contribution to the VHE emission of the external inverse Compton (EIC) process on this radiation field ({\it in red}). It appears to be comparable to the SSC component. Further studies of simultaneous variabilities in the X-ray and VHE domains with CTA would allow to disentangle between SSC and EIC.

\psfrag{theta}{$\theta$}
\psfrag{vj}{$v_j$}
\psfrag{jet axis}{jet axis}
\psfrag{rb}{$r_b$}
\psfrag{Gamma_b}{$\Gamma_b$}

Regardless of the origin of the inverse Compton bump, we found out that Cen\,A could be detectable by H.E.S.S. within 50\,h of observation, and should be easily reachable with the next generation of \v{C}erenkov instruments as CTA.

\section{Conclusion}

Our scenario shows the ability to extend the interpretation of TeV blazars by leptonic models to active galactic nuclei with intermediate beaming which jet is not strictly aligned with the line of sight.

The {\it GLAST} mission and H.E.S.S.\,II project, and CTA in a near future, will allow to probe the sub-TeV spectral region, which is still poorly known and yet essential to constrain the inverse Compton bump. This is required to distinguish between leptonic or hadronic origin of the high energy emission of the jet in these objects.

%\section*{Acknowledgments}
%This is where one places acknowledgments for funding bodies etc.
%Note that there are no section numbers for the Acknowledgments, Appendix
%or References.

\end{document}